\documentclass{emulateapj}
\pdfoutput=1
\usepackage{natbib}
\usepackage{times}

\begin{document}

\shorttitle{Supernova Neutrinos and Shock Breakout}
\shortauthors{Kistler, Haxton, \& Y{\"u}ksel}

\title{Tomography of Massive Stars from Core Collapse to Supernova Shock Breakout}

\author{Matthew D.~Kistler\altaffilmark{1,3},
W.~C.~Haxton\altaffilmark{1}, and
Hasan Y{\"u}ksel\altaffilmark{2}
}

\affil{$^1$Lawrence Berkeley National Laboratory and Department of Physics, University of California, Berkeley, CA 94720, USA}
\affil{$^2$Theoretical Division, MS B285, Los Alamos National Laboratory, Los Alamos, NM 87545, USA}
\altaffiltext{3}{Einstein Fellow}


\begin{abstract}
Neutrinos and gravitational waves are the only direct probes of the inner dynamics of a stellar core collapse.  They are also the first signals to arrive from a supernova and, if detected, establish the moment when the shock wave is formed that unbinds the stellar envelope and later initiates the optical display upon reaching the stellar surface with a burst of UV and X-ray photons, the shock breakout (SBO).  We discuss how neutrino observations can be used to trigger searches to detect the elusive SBO event.  Observation of the SBO would provide several important constraints on progenitor structure and the explosion, including the shock propagation time (the duration between the neutrino burst and SBO), an observable that is important in distinguishing progenitor types.  Our estimates suggest that next generation neutrino detectors could exploit the overdensity of nearby SNe to provide several such triggers per decade, more than an order of magnitude improvement over the present.
\end{abstract}

\maketitle

\section{Introduction}
\label{section:introduction}
Core-collapse supernova observations have improved rapidly in quantity and quality over the past decade, and with the coming large synoptic surveys, rates should soon advance by an additional two orders of magnitude or more, with $\sim\,$10$^5$ events per year likely to be found within $z$$\,\sim\,$0.5--1 \citep{Lien09}.  With the new data has come a greater appreciation of the variety of stellar evolutionary phenomena leading to these events.  These include a wide range of possibilities associated with the loss/modification of the envelope.  For very massive stars, these include strong outbursts occurring within a few years prior to the collapse \citep{Pastorello:2007mv,Smith2012,Ofek:2013mea,Fraser:2013wva} and circumstellar interactions during the SN \citep{GalYam2012}, while premortem enshrouding by dust has been seen in lower-mass stars near the core-collapse threshold \citep{Prieto:2008bw,Thompson:2008sv}.  The prevalence of binary evolution suggests that mass transfer could play a role prior to a substantial fraction of collapses \citep{Sana:2012px}.  However, we have not found signatures of impending core collapse that are useful observationally -- that would allow us to find and study stars within a designated time before collapse.

A successful core-collapse supernova requires the ejection of the remaining stellar envelope by an outward propagating shock wave that originates deep within the star's core, where it can be boosted initially by neutrino heating of the matter behind the shock front.  The radiative precursor -- the first radiation to ``leak out" ahead of the shock front, marking the beginning of shock breakout (SBO), a burst of UV/soft X-ray photons -- appears when the shock front reaches a point in the star's envelope where
the  optical depth $\tau$$\,\sim\,$25$\,\sim\,$$c/v_\mathrm{shock}$ \citep{Colgate68,Falk1977,Klein1978}.  The SBO grows in luminosity, peaking at $\tau$$\,\sim\,$1 and producing an integrated energy release in radiation of $\sim\,$$10^{46}$--$\,10^{49}\,$erg (e.g., \citealt{Ensman(1992)}).

The SBO contains a great deal of information about the exploding star (e.g., \citealt{Matzner(1999)}).  However, obtaining quality data from such events is exceedingly difficult, since the SBO is the first detectable photon signal to arrive from a SN and has a duration of only seconds to hours (depending on the progenitor), much less than the time that a SN is bright optically.  Further, Earth's atmosphere is opaque to the SBO spectral peak, and space telescopes cannot typically observe more than one nearby galaxy at a time.

To observe a SBO at present requires either observations of deep fields or a great deal of luck.  Observations of distant SNe with limited sampling of light curves can be open to interpretation, such as a SN at $z$$\,=\,$0.185 detected by GALEX, for which \citet{Schawinski:2008ba} claimed that the SBO was seen, while \citet{Gezari:2008jb} concluded otherwise.  The detection of a soft X-ray burst preceding SN~2008D, interpreted as a SBO from a dense Wolf-Rayet wind by \citet{Soderberg:2008uh}, was due to serendipitous {\it Swift} observations of the host galaxy as the star exploded, an exceptional circumstance.

We discuss the possibility of mitigating these difficulties by exploiting the two early signals of core collapse available to observers, neutrinos and gravitational waves \citep{Marek:2008qi,Yakunin:2010fn,Ott:2012kr}, concentrating on the former.  We discuss two strategies, each with positive and negative attributes.  The first, a Galactic SN, would produce an unambiguous signal in neutrino detectors today (see \citealt{Ikeda:2007sa,Abbasi:2011ss,Scholberg:2012id}) that could be combined with shock breakout data to constrain properties of the collapsing star, but suffers from an unfavorable event rate of  $\sim\,$1 per century. 

\begin{figure}[t!]
\includegraphics[width=3.4in,clip=true]{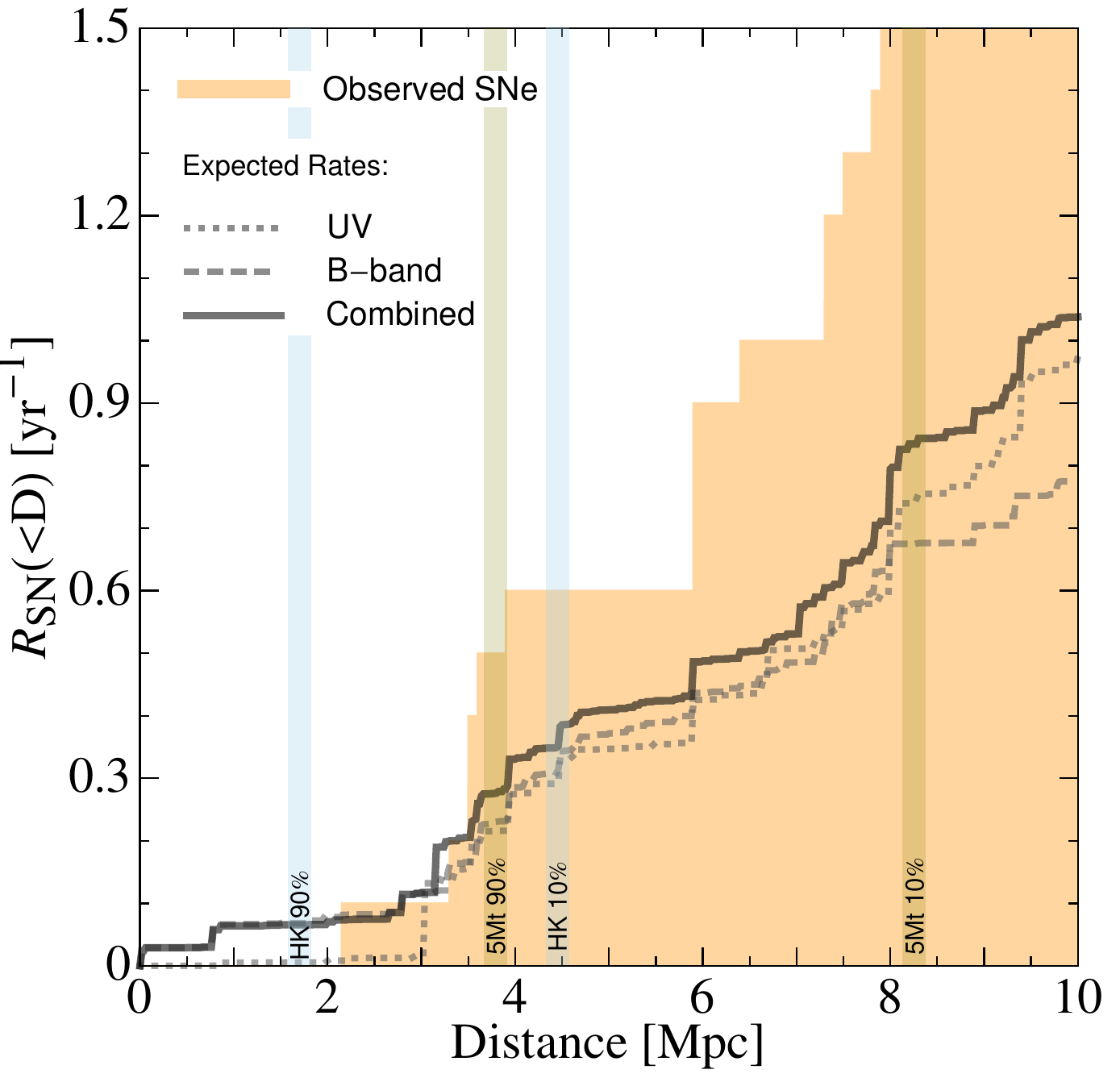}
\caption{The cumulative rate of nearby core-collapse SNe versus distance.  Shown are the expectations based on UV emission from galaxies in \citet{Lee:2009by} ({\it dotted}), $B$-band emission from the \citet{Karachentsev:2004dx} catalog ({\it dashed}), and the 589 galaxies in our combined set ({\it solid}), compared to the 21 SNe observed in 1999-2008 as compiled in \citet{Kistler:2008us} ({\it steps}).  We denote nominal reaches of 560~kton (Hyper-K) and 5~Mton neutrino detectors using the distances at which 90\% or 10\% of SNe are detectable (see \citealt{Kistler:2008us}; \citealt{Ando:2005ka} for probability curves).\\
}
\label{cumult}
\end{figure}

The second, and our main focus here, is SNe within nearby galaxies.  A number of studies have confirmed the feasibility of water Cherenkov detectors in the $\sim\,$0.5--1 megaton range, and serious efforts to proceed with such neutrino detectors within a decade have been made around the world \citep{Abe:2011ts,Goon:2012if,Autiero:2007zj}.  As these will have a reach of few Mpc (see Fig.~\ref{cumult}), their SN detection rates can exceed one per decade.  A multi-Mton instrument would increase this to nearly one event per year.  Two techniques to reach a $\sim\,$5~Mton mass have been put forward.  Both employ strategies to provide the overburden required to adequately suppress cosmogenic backgrounds, without excavating a mountain of rock.  The DeepTITAND concept, addressed in \citet{Suzuki:2001rb} and \citet{Kistler:2008us}, is based on large tanks of pure water submerged beneath the sea, while the \citet{IceCube:2011ac} has discussed a densely-instrumented array in South Pole ice within IceCube.

Detection of both the neutrino burst and the SBO would determine the shock propagation time, placing an important new constraint on progenitor star properties.  Even if directional information from the neutrinos is insufficient to isolate the source -- which would likely be the case -- we find that the candidate SN host galaxies would be limited in number, so that targeted campaigns could be launched to discover the UV/X-ray burst from the SBO using conventional methods.  We also discuss searches for below-threshold neutrino counts in the event that non-triggered SBO surveys prove successful, the combination of which would achieve the same end.  The distance range covered by this technique also coincides well with surveys that are imaging prospective SN progenitors.  Connecting the SN/SBO event with pre-explosion optical images would further improve our characterization of the star, with important implications for better understanding SN morphologies.

\section{Break on Through}
Observational efforts to measure the progression of the SN shock wave from the stellar core to surface will require neutrino or gravitational wave measurements of the collapse in conjunction with techniques to capture the SBO signal that follows after a delay that varies from star to star.  As motivation, we first discuss the relationship of the propagation time to the SBO observables addressed by \citet{Matzner(1999)}.  There are four dimension-full scales that arise when dealing with an adiabatic, radiation-dominated, and non-gravitating explosion: the explosion energy $E_\mathrm{in}$, the core mass $M_\mathrm{NS}$, the ejected mass $M_\mathrm{ej}$\,$=$\,$M_*$\,$-$\,$M_\mathrm{NS}$, and the progenitor star's radius $R_*$.   Here $M_*$\,$=$\,$ M_\mathrm{ej} $\,$+$\,$ M_\mathrm{NS}$ is the total mass.  We describe below the functional dependences of SBO observables on these four scales in simple polytrope models.  

\citet{Matzner(1999)} discussed three observables connected with shock emergence: the post-shock radiation temperature $T_\mathrm{se}$, the SBO energy $E_\mathrm{se}$, and the timescale for radiation to diffuse out of the shock $t_\mathrm{se}$ (for non-relativistic ejecta).  These were evaluated in the context of an analytical model, a progenitor whose envelope can be represented as a polytrope.  They provided results for $n$\,$=$\,$3/2$ and $n$\,$=$\,$3$ polytropes, with the latter yielding
\begin{eqnarray}
T_\mathrm{se} &=& 1.31 \times 10^6\,\mathrm{K}~\left( \frac{\kappa}{0.34\, \mathrm{cm^2~g^{-1}}} \right)^{-0.14}  \left( \frac{\rho_1}{\rho_*} \right)^{0.046} \nonumber \\
&\times& \left( \frac{E_\mathrm{in}}{10^{51}\,\mathrm{erg}} \right)^{0.18} \left( \frac{M_\mathrm{ej}}{10\,M_\odot} \right)^{-0.068}  \left( \frac{R_*}{50\, R_\odot} \right)^{-0.48} \nonumber
\end{eqnarray}
\begin{eqnarray}
E_\mathrm{se} &=& 7.6 \times 10^{46}\mathrm{erg}~\left( \frac{\kappa}{0.34\, \mathrm{cm^2~g^{-1}}} \right)^{-0.84}  \left( \frac{ \rho_1}{ \rho_*} \right)^{-0.054} \nonumber \\
&\times& \left( \frac{E_\mathrm{in}}{10^{51}\, \mathrm{erg}} \right)^{0.58} \left( \frac{M_\mathrm{ej}}{ 10\,M_\odot} \right)^{-0.42}  \left( \frac{R_*}{ 50\, R_\odot} \right)^{1.68} \nonumber
\end{eqnarray}
\begin{eqnarray}
t_\mathrm{se} &=& 40\,\mathrm{s}~\left( \frac{\kappa}{ 0.34\, \mathrm{cm^2~g^{-1}}} \right)^{-0.45}  \left( \frac{ \rho_1}{ \rho_*} \right)^{-0.18} \nonumber \\
&\times& \left( \frac{E_\mathrm{in} }{10^{51}\,\mathrm{erg}} \right)^{-0.72} \left( \frac{M_\mathrm{ej} }{ 10\,M_\odot} \right)^{0.27}  \left( \frac{R_*}{ 50\, R_\odot} \right)^{1.90} .
\label{equc1}
\end{eqnarray}
Here $\kappa$ is the opacity and the density ratio $\rho_1/\rho_*$ can be related to the ratio of $M_*/M_\mathrm{ej}$ by a coefficient that depends on the polytrope, e.g.,
\begin{equation}
 \frac{\rho_1}{ \rho_*}  \equiv  \frac{R_*^3}{ M_\mathrm{ej}}   \left( \frac{G M_*}{ (n+1)\, K\, R_*} \right)^n
~ \stackrel{n=3}{\longrightarrow}~ 0.324\, \frac{M_*}{ M_\mathrm{ej}},
\end{equation}
where $K$ is the equation-of-state parameter, $p_0$\,$=$\,$K\, \rho_0^{(n+1)/n}$.  One motivation for the work of Matzner and McKee was to relate observables connected with the SBO and expanding ejecta with parameters characterizing the progenitor's structure and the explosion, $M_\mathrm{ej}$, $R_*$, and $E_\mathrm{in}$.

The present study of the shock wave propagation time, defined here as
\begin{equation}
	\Delta t = t_{\rm SBO}-t_{\rm CC} = \int_{R_{\rm NS}}^{R_*} \frac{dr}{v_s(r)}\,,
	\label{equ1}
\end{equation}
has a similar motivation.  An analytic form for the shock velocity $v_s(r)$ that accounts for acceleration due to the sweeping up of enclosed envelope mass, $m(r)$, and the declining density gradient, $\rho_0(r)$, was given by \citet{Matzner(1999)} (who showed the formula nicely reproduces numerical velocity profiles, as can be seen from Fig.~2 of their paper).  With this substitution in Eq.~(\ref{equ1}), $\Delta t$ becomes
\begin{equation}
  \Delta t = 1.26  \int_{R_{\rm NS}}^{R_*} dr\,\left[\frac{m(r)}{E_\mathrm{in}}\right]^{1/2}  \left[\frac{\rho_0(r)\,r^3}{m(r)}\right]^{0.19}. 
	\label{equ2}
\end{equation}
While we will evaluate this for specific progenitors below, we can make contact with \citet{Matzner(1999)} by calculating $\Delta t$ for polytropes, to illustrate its functional dependence on progenitor properties.  In \citet{Matzner(1999)} the polytrope could be restricted to the envelope, in which case the Lane-Emden equation reduces to a first-order equation with a boundary condition at the inner surface of the envelope.  In contrast, as $\Delta t$ involves an integral over most of the star, here a polytrope is a more aggressive approximation.  We find for $n=3$ and $3/2$
\begin{eqnarray}
  \Delta t_{n=3} = 6851\,s~\left( \frac{10^{51}\,\mathrm{erg}}{ E_\mathrm{in}} \right)^{1/2} \left( \frac{M_\mathrm{ej} }{10\,M_\odot} \right)^{1/2} \left( \frac{ R_*}{ 50\,R_\odot} \right) \nonumber \\
\times \left[1 -0.407 \left( \frac{M_\mathrm{NS} }{ M_\mathrm{ej}} \right)^{0.81} +0.285 \left(\frac{M_\mathrm{NS} }{ M_\mathrm{ej}} \right)^{1.12} \right]~~~~~ \nonumber
\end{eqnarray}
\begin{eqnarray}
  \Delta t_{n=\frac{3}{2}} = 7226\,s~\left( \frac{10^{51}\,\mathrm{erg} }{ E_\mathrm{in}} \right)^{1/2} \left( \frac{M_\mathrm{ej}}{10\,M_\odot} \right)^{1/2} \left( \frac{ R_*}{ 50\,R_\odot} \right) \nonumber \\
\times \left[1 -0.738 \left( \frac{M_\mathrm{NS} }{ M_\mathrm{ej}} \right)^{0.80} +0.467 \left(\frac{M_\mathrm{NS} }{ M_\mathrm{ej}}
\right)^{1.20} \right].~~~~
\end{eqnarray}
These two results are very similar, which perhaps suggests relatively little sensitivity to polytrope index or variations of that index that might be appropriate for different portions of the star.  The factors in the square brackets are fits to numerical results and are very accurate representations of the dependence of Eq.~(\ref{equ2}) on $R_\mathrm{NS}$.  Deep interiors of core collapse supernovae are often described as $n$\,$=$\,$3$ polytropes, as this corresponds to a relativistic electron gas equation of state.  As discussed in \citet{Matzner(1999)}, $n$\,$=$\,$3$ is appropriate for the radiative envelopes of blue supergiants, but a better fit to models for the inner 75\% of the mantle (by mass) is obtained with an effective index $n$$\, \sim \,$$2.1-2.4$.

\begin{figure}[t]
\includegraphics[width=3.4in,clip=true]{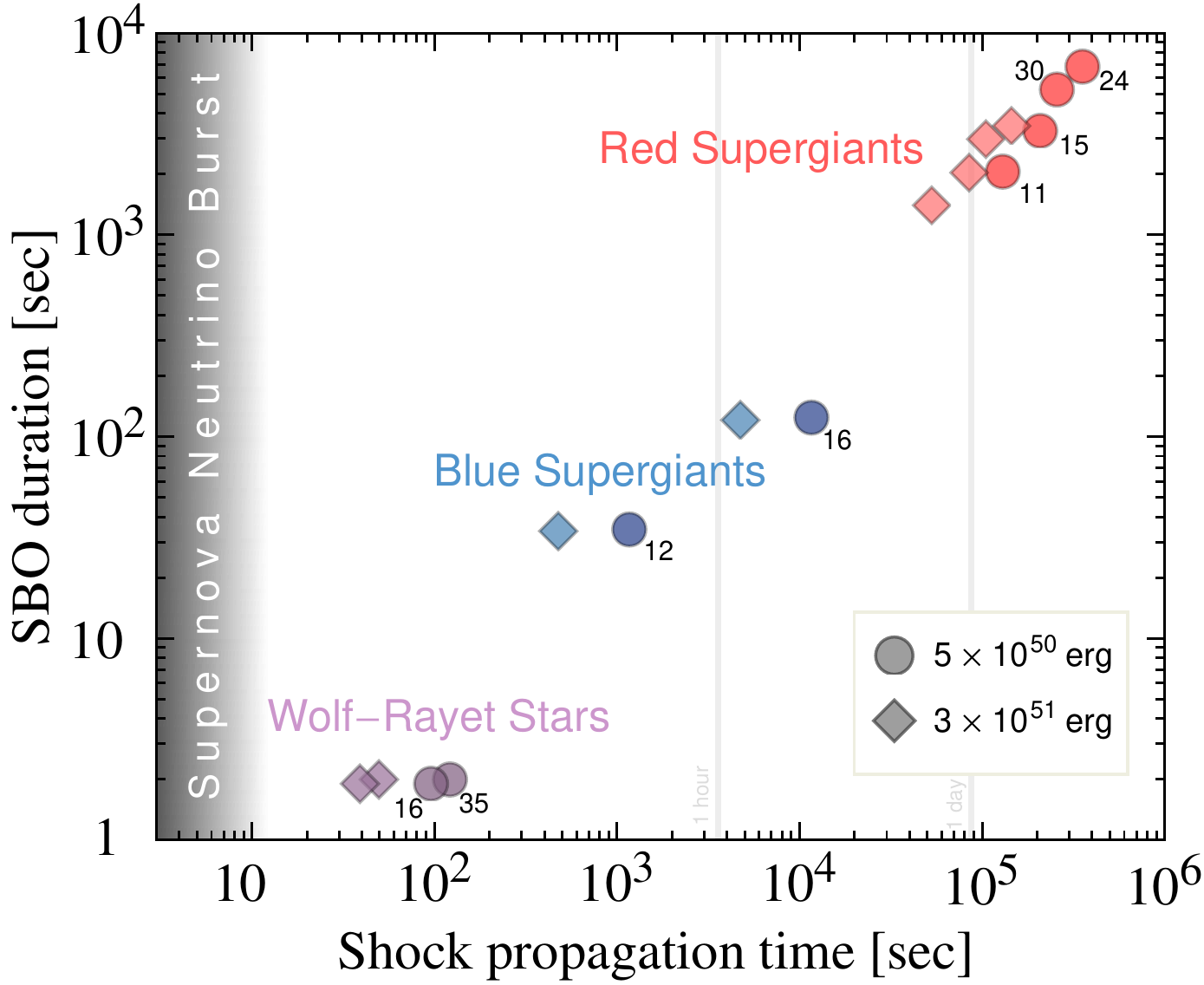}
\caption{A comparison of shock breakout (SBO) durations versus shock propagation times in the envelopes of SN progenitor models, as calculated for a variety of initial masses from $11-35\,M_\odot$ (as labeled), using density profiles from \citet{Woosley:2002zz} for RSG and \citet{Woosley:2005gy} for BSG and Wolf-Rayet stars, with shock energies of 0.5$\times$ and 3$\times\,$$10^{51}\,$erg.\\}
\label{shockprop}
\end{figure}

We have also evaluated Eq.~(\ref{equ2}) directly from progenitor models.  Fig.~\ref{shockprop} shows the resulting propagation times for a 11--30$\,M_\odot$ range of non-rotating red supergiant (RSG) progenitors from \citet{Woosley:2002zz}, as well as 12 \& 16$\,M_\odot$ blue supergiant (BSG) and 16 \& 35$\,M_\odot$ Wolf-Rayet star models from \citet{Woosley:2005gy}, with the inner $1.4\,M_\odot$ forming a neutron star.  We have used $E_\mathrm{in}$$\,=\,$0.5$\times$ and 3$\times$$10^{51}\,$erg to bracket the canonical $10^{51}\,$erg.  The polytrope results yield propagation times that generally agree to within $\sim\,$10--20\%.

The convective RSG envelopes extend up to $\sim\,$$1500\,R_\odot$, while BSG radii are typically limited to $\sim\,$$25\,R_\odot$.  From the polytrope results one expects propagation times to be proportional to radii, and Fig.~\ref{shockprop} shows the expected gap of $\sim 50$ between RSG and BSG times.  In Wolf-Rayet stars, thought to give rise to Type Ib/Ic SNe ($\sim\,$10--20\% of all SNe), the shock arrives at the surface very quickly, as the strong winds in such stars lead to complete loss of their envelopes.  As the SBO timescale is comparable to that of neutrino emission, little early warning would be available, although an optically-thick wind may delay emergence (e.g., \citealt{Balberg:2011qt,Chevalier:2012zf}).

While observing the SBO spectrum and light curve would provide several pieces useful for a forensic study of the departed star (see \citealt{Calzavara:2003ap}), timing alone would suggest a type and mass.  An illustrative example of the utility of this feature is in settling disagreements in mass estimates provided by hydrodynamic SN modeling and pre-SN progenitor imaging.  SN~2004et, for instance, originated in NGC~6946, within range of a 5~Mton detector.  As in Fig.~\ref{shockprop}, a $9_{-1}^{+5}\,M_\odot$ progenitor \citep{Smartt:2008zd} gives an expected propagation time of $\lesssim\,$1~day, while for a $27\pm 2\,M_\odot$ model \citep{Utrobin:2011} this would be $\sim\,$1~day later.

\section{The Harbinger}
The critical aspect of a triggered search for SBO is the ability to detect that a massive-stellar core has collapsed before the SBO photons arrive\footnote{The earliest photons actually arise from heating of the hydrogen envelope by the core $\bar{\nu}_e$ burst.  However, for a low-mass Fe-core supernova the energy deposition is $\sim (1-3) \times 10^{43}$ ergs, depending on assumptions about neutrino temperatures and oscillation effects, far below the SBO energies given above.  Specific gamma-ray signals associated with $\bar{\nu}_e + p \rightarrow n + e^+$ -- from $e^+e^-$ annihilation and from $n+p \rightarrow d + \gamma$ -- have also been considered, but found to be undetectable with present or envisioned detectors \citep{Qian07}.} (with a low rate of false positives) and to notify the astronomical community rapidly.  Neutrinos are of particular interest since the data from SN~1987A \citep{Bionta:1987qt,Hirata:1987hu} provide the general SN neutrino burst properties.  Much of the existing literature on this topic is useful (e.g., \citealt{Scholberg:1999tm,Calzavara:2003ap}), yet somewhat dated, based on those detectors that were in operation a decade ago.  Here we will consider both Galactic and extragalactic events in turn.\\

\subsection{One in Every Crowd: Locating a Galactic SN}

The task of identifying the SBO can be simplified to the extent that data from the neutrino harbinger can be quickly analyzed to pinpoint the angular region likely to contain the SN.  This is essential in the case of a Galactic event, as otherwise the scanning must encompass all angles.  There has been some important work preparing the community for rapid sharing and coordinated analysis of data, with the organization of the SuperNova Early Warning System (SNEWS; \citealt{SNEWS}) being an outstanding example.  We first review the conclusions for a Galactic SN, the focus of previous work.

Studies have focused on two techniques, exploiting the pointing capabilities of an individual water, heavy water, or liquid argon neutrino detector to locate an event, or coordinating the results from several detectors to determine location by timing and triangulation (gravitational waves also rely on triangulation and with multiple detectors, possibly including the use of templates, might achieve a localization comparable to neutrinos for a Galactic SN, e.g., \citealt{Aasi:2013wya,Wen:2010cr}, although our focus shall be on neutrinos).  The consensus from past studies has been that the more sensitive technique is pointing \citep{Scholberg:2012id}.

In a water detector, the limiting factor is the finite angular resolution in reconstructing the neutrino direction via the Cherenkov light from the resulting electron/positron.  Despite the much weaker signal for $\nu_e$$\,+\,$$e^-$$\, \rightarrow \,$$\nu_e$$\,+\,$$e^-$ elastic scattering -- typically $\sim\,$1/30th the SN $\bar{\nu}_e$$\, + \,$$p$$\, \rightarrow \,$$n$$\, + \,$$e^+$ signal, depending on the detector threshold -- this is the more effective pointing reaction due to the sharply forward-peaked cross section.  (The charged current $\bar{\nu}_e$ reaction in water is approximately flat in angle, requiring an analysis that exploits small, energy-dependent anisotropies in the detected $\bar{\nu}_e$ events.)

Super-Kamiokande will record $\sim\,$300 $\nu_e$$\,+\,$$e^-$$\, \rightarrow \,$$\nu_e$$\,+\,$$e^-$ events from a $\sim\,$10~kpc Galactic SN, leading to an excess in the SN direction on top of the nearly-isotropic background of $\sim\,$$10^4$ $\bar{\nu}_e$$\, + \,$$p$$\, \rightarrow \,$$n$$\, + \,$$e^+$ events \citep{Ikeda:2007sa}.  \citet{Beacom:1998fj} concluded that the localization possible in this case is an angular cone of $\sim\,$5$^\circ$ radius.  The Collaboration is considering adding gadolinium to the detector to allow for the detection of neutrons, which are currently invisible in Super-K \citep{Beacom:2003nk}.  If this is done, the pointing capabilities could be improved by vetoing $\bar{\nu}_e$ events to isolate the otherwise-indistinguishable $e^-$ scattering.  The future detector Hyper-Kamiokande \citep{Abe:2011ts}, with a fiducial volume 25 times that of Super-K, could localize a Galactic event to $\sim\,$1$^\circ$.

In their exploration of the potential of timing and triangulation, \citet{Beacom:1998fj} exploited the $\tau\,$$\sim\,$30~ms rise time in the neutrino signal, determining that the event could be localized to within a time that scaled as $\tau/\sqrt{N}$, where $N$ is the number of detected events.  Their conclusion, that triangulation was a factor of $\sim\,$100 less effective in limiting $\delta(\cos{\theta})$, when compared to pointing by elastic scattering, was based on a comparison of the Super-Kamiokande and SNO detectors, and thus was limited by the smaller detector.  It is possible that this is too pessimistic, however, as there may be other, more distinctive time variations in the SN neutrino flux.

Recent studies have seen high-frequency (10--200~Hz) stochastic variations in 2D and 3D simulations of the SN flux -- a consequence of convective instabilities and the standing accretion shock instability -- and concluded that these variations would currently be detectable in IceCube for a SN within $\sim\,$2--10 kpc, depending on assumptions, despite a very limiting signal/noise ratio in that detector of $\sim\,$1/10 \citep{Lund,Brandt,Tamborra:2013laa}.  The SN neutrino event rate in Hyper-Kamiokande during the first several seconds of the burst should be $\sim\,$50/ms, about 1/3 the rate in IceCube, but with negligible background.  Thus, once more than one detector of this class is operational, it is possible that timing via detailed pattern matching would greatly increase the power of this technique.

\subsection{Supernovae in the Local Universe}

While future neutrino advances will localize a Galactic SN to within a degree, it will not solve the problem that motivates this study, the $\sim\,$hundred-year wait between events.  Further, whenever the next Galactic SN does occur, it will still only represent one class of the great variety of SNe.  Unfortunately, neither of the techniques described above can be applied successfully to the extragalactic SNe of interest here, as both depend fundamentally on large statistics.

The prospect of detecting a SN core collapse from far beyond the Galaxy long appeared to be impractical.  However, in \citet{Ando:2005ka} it was recognized that a 1~Mton detector would expect $\sim\,$1 $\bar{\nu}_e$$\, + \,$$p$$\, \rightarrow \,$$n$$\, + \,$$e^+$ event from out to a few Mpc, which could be searched for after an optical discovery.  A SN in Andromeda, which is also an infrequent source of SNe, would produce $\sim\,$40 events in Hyper-Kamiokande.  If we set as our goal reaching starburst galaxies such as M82 and IC342, where SNe are far more frequent, then the best we can hope to achieve is a signal distinguishable from background.  As these starburst galaxies are at distances of over 3 Mpc, each core collapse would produce $\sim\,$10 counts/event in a 5~Mton detector and $\sim\,$2--3 in Hyper-Kamiokande.

We emphasize that, for the purpose of triggering, an independent detection of the core collapse is essential.  Following the methods in \citet{Kistler:2008us}, we address the prospects for detecting ``mini-bursts'' of core-collapse neutrino events.  For a 5~Mton detector, background rates dictate that $\ge\,$3 SN $\bar{\nu}_e$ events are required to reduce false triggers.  In a 560~kton detector, such as Hyper-Kamiokande, the absolute background rate is lower and adding gadolinium should allow a reduction to $\ge\,$2 events, with a lower energy threshold (we use $E_{\bar{\nu}_e}$$\,\gtrsim\,$16~MeV).  In Fig.~\ref{cumult}, we indicate nominal detector reaches by the distances at which 90\% or 10\% of SNe are detectable for these thresholds (complete probability curves are available in \citealt{Kistler:2008us} and \citealt{Ando:2005ka}).

\begin{figure}[t!]
\includegraphics[width=3.42in,clip=true]{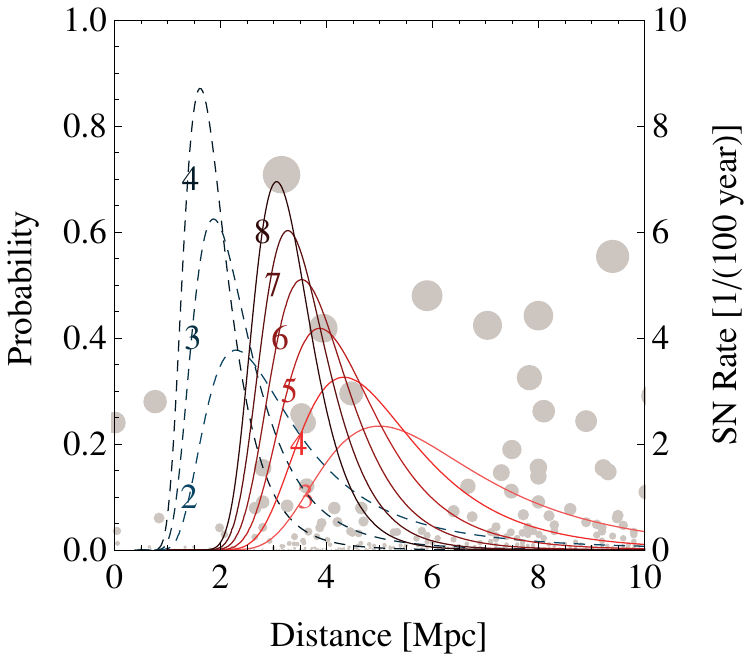}
\caption{The probability of a core-collapse originating at an unknown distance based {\it only} on a given number of neutrino events detected in 560~kton ({\it dashed}) and 5~Mton water Cherenkov detectors ({\it solid}; as numbered).  Also shown are the supernova rates for galaxies used in our estimate of the local rate, shown as {\it circles} (with area proportion to the rate) at the distance of each galaxy (note that these do not factor into the curves displayed, but are reflected in Table~\ref{tab:params}).
}
\label{prob}
\end{figure}

To estimate the nearby rate of burst detections, we use SN rates derived from two techniques.  We consider $B$-band emission, using the updated empirical conversion factors from \citet{Li:2010iii}, from the 451 galaxies in the \citet{Karachentsev:2004dx} catalog, as well as the $B$-band and UV-based rates (rather than H$\alpha$; see \citealt{Botticella:2011nd}) from the 315 galaxies in \citet{Lee:2009by} assuming a minimum SN mass of $8\,M_\odot$.  To reach the greater completeness needed for neutrino burst association, we merge the two catalogs, using the average of the $B$-band and UV rate for galaxies appearing in both, for a total of 589 galaxies within $\sim\,$$11\,$Mpc (see Fig.~\ref{prob}).

In Fig.~\ref{cumult}, we see that the galaxy catalog rate falls below the observations of nearby SNe.  As in \citet{Kistler:2008us}, we assume a generic Fermi-Dirac SN $\bar{\nu}_e$ spectrum with an average energy of 15 MeV yielding a total energy of $5\times 10^{52}$~erg.  Using the catalog (excluding the Milky Way) and the observed SNe as lower and upper bounds, we find expected decadal core-collapse detection rates of $\sim\,$4--8 in 5~Mton and $\sim\,$1--2 in 560~kton fiducial volumes.  Both are well above current capabilities, set by the Galactic rate of one per 30--100~yr.

The number of potential SN host galaxies within neutrino range of Earth is finite, but if the only information available is that a neutrino burst has been detected, all would need to be scanned to find the SBO.  Here we argue that the specific details of the neutrino burst  -- the number of counts, and possibly the hardness of the neutrino spectrum -- and the size of the source -- e.g., M82 has an angular size of less than 0.5$^\circ$ -- will allow one to focus telescope time on a subset of candidate galaxies that reside at an appropriate distance from Earth.

\begin{figure}[t!]
\includegraphics[width=3.2in,clip=true]{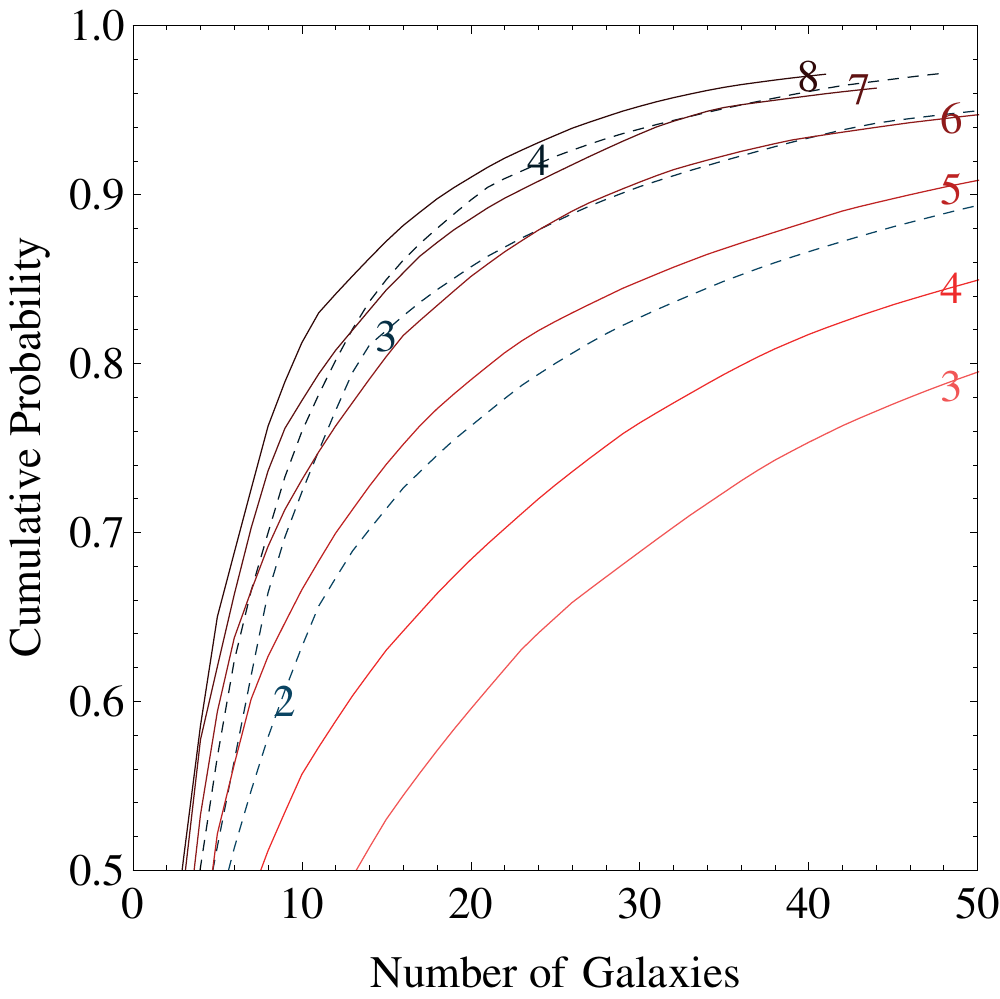}
\caption{Cumulative probability versus number of galaxies, with curves corresponding to 560~kton ({\it dashed}) and 5~Mton detectors ({\it solid}) as in Fig.~\ref{prob}.
}
\label{cumprob}
\end{figure}

We next examine the quantitative prospects for deriving a distance to the core collapse from neutrinos alone.  Fig.~\ref{prob} displays the probability of a SN arising from an unknown distance given only the number of $\bar{\nu}_e$ events seen in a 560~kton or 5~Mton detector (assuming our fiducial $\bar{\nu}_e$ spectrum, an input that we expect future simulations will further refine).  Even for low event counts, this narrows the range of distances that needs to be considered, so that a SBO search can be confined roughly to a shell of neighboring galaxies.

Variations in $\bar{\nu}_e$ production because of variations in the progenitor can be estimated from the library of progenitors recently generated by \citet{Nakazato}.  Detection rates in water Cherenkov detectors are approximately proportional to the number of $\bar{\nu}_e$ emitted times $\langle E_{\bar{\nu}_e} \rangle^2$ (due to the energy-dependence of the inverse-beta cross section).  Forming this figure-of-merit from the various metallicity $Z$$\,=\,$0.02 progenitors in Table~1 of \citet{Nakazato} and demanding that our associated uncertainty encompasses the full range, we find that event rates can vary by $\pm$ 42\% (although one should account for a declining stellar initial mass function).  The Poisson uncertainty on the distance inherent in Fig.~\ref{prob} should tend to be more important for small numbers of counts.  The effects of neutrino oscillations have also been evaluated recently in connection with the diffuse supernova neutrino background \citep{Cecilia} and shown to be relatively modest.

Taking into account both the probabilities and SN rates shown in Fig.~\ref{prob}, we determine the overall probability, $P_{n}$,  for a SN to have arisen in each galaxy in our set given an observed number of neutrino counts, $n$, in a 560~kton or 5~Mton detector.  In Table~\ref{tab:params}, we rank these to show the most likely locations for some of the most likely $\bar{\nu}_e$ event counts.  Although we consider 589 galaxies, we see that strongly star-forming galaxies are the most probable and limited in number.  Fig.~\ref{cumprob} displays how larger event counts limit the number of galaxies required to reach a given probability when using this set.

\begin{table*}[t]
\caption{Estimated probability, $P_{n}$, for a galaxy to host a SN given $n$ detected $\bar{\nu}_e\,p\rightarrow e^+\,n$ events in a 560~kton or 5~Mton detector.}
\label{tab:params}
\begin{ruledtabular}
\begin{tabular}{cccccc}
$P_{2,~(560~{\rm kton})}$ & $P_{3,~(560~{\rm kton})}$ & $P_{3,~(5~{\rm Mton})}$ &  $P_{4,~(5~{\rm Mton})}$ & $P_{6,~(5~{\rm Mton})}$ &  $P_{8,~(5~{\rm Mton})}$  \\ \hline
IC~342 (0.220)			& IC~342 (0.258)		& NGC~6946 (0.093)	& NGC~253 (0.11)		& IC~342 (0.218)		&  IC~342 (0.333)\\
NGC~253 (0.078)		& Maffei~2 (0.087)		&NGC~253 (0.063)		& M83 (0.086)			& NGC~253 (0.134)		& M82 (0.091)\\
M82 (0.063)			& M82 (0.059)			&M83 (0.062)			& IC~342 (0.076)		& M82 (0.093)			& NGC~253 (0.084)\\
Maffei~2 (0.058)		& NGC~253 (0.059)		&M101 (0.054)			& NGC~6946 (0.075)	& NGC~4945 (0.087)	& NGC~4945 (0.078)\\
NGC~4945 (0.056)		& NGC~4945 (0.052)	&M51 (0.037)			& NGC~4945 (0.052)	& M83 (0.061)			& Maffei~2 (0.065)\\
\end{tabular}
\end{ruledtabular}
\end{table*}

This strategy for limiting the SBO search is compatible with the anticipated meager number of events that will be detected from an extragalactic SN.  For example, the detection of 20 events would determine the flux to $\sim$ 22\%, well below the 42\% theoretical flux uncertainty we associated with the unknown properties of the SN progenitor.  A more significant difference could arise if collapse proceeds to a black hole without a robust explosion.  In this case, the flux from the short-lived proto-neutron star should be larger with a harder spectrum (\citealt{Burrows:1988ba,Baumgarte:1996iu,Sumiyoshi:2007pp,Fischer2009,Nakazato}; see, e.g., Figure~2 of \citealt{Yuksel:2012zy}).  While we have assumed this population to not be substantial, the fraction of core collapses that end this way is not well constrained, with a current limit of $\lesssim\,$50\% \citep{Kochanek:2008mp,Horiuchi:2011zz}.  If this fraction is in the tens of percent range, it would start to affect the strategy of neutrino searches, since these events could be detectable from greater distances.  This itself would be quite a significant discovery.  In this case their presence should be first noticed via their imprint on the spectrum of the cosmic MeV neutrino background, as discussed in \citet{Yuksel:2012zy} (see also \citealt{Lunardini:2009ya}), or possibly a new class of optical transients that we discuss later, so that the true rate can be estimated independently.

We believe that the analysis presented here could be further refined, to take into account additional information from the detected neutrino events:  for example, as progenitors that release more energy in $\bar{\nu}_e$ tend to produce harder spectra, an improved distance indicator could be designed that incorporates information on $\langle E_{\bar{\nu}_e} \rangle$.  But even with such elaborations, the basic idea of the analysis is so simple that a collaboration like Hyper-Kamiokande should be able to make a quick determination of the range of likely distances to the progenitor, thereby informing SBO searches.

\section{The Chase}
Along with the shock propagation time, the duration of the SBO event determines the requirements for a triggered search.  This is set by a combination of the photon diffusion time, $t_d$, and the light travel time over the stellar surface, $t_*$$\, \sim \,$$r_* / c$.  Following \citet{Calzavara:2003ap}, we set $t_d$$\,=\,$$l_s/v_s$ based on the point where $\tau_{\rm shock}$$\, = \,$$c/v_s$$\, = \,$$\tau_*$$\,=\,$$\int_{l_s}^{r_*}dr\, \kappa \, \rho$, with $\kappa$$\,=\,$$0.34\,$cm$^2 \,$g$^{-1}$, to find $t$$\, \sim \,$$(t_d^2 + t_*^2)^{1/2}$.  The results for the progenitor models are shown in Fig.~\ref{shockprop}.  For a fixed stellar mass/radius, the polytrope solutions tend to yield somewhat longer $t_d$ values, although this depends on the details of the outer density profile at the time of explosion and should thus be accounted for in future modeling.  

Ideally one would like to accurately determine both the duration of the SBO and the time of shock breakout relative to a harbinger of core collapse.  These two times are in fact correlated, though they differ in their detailed dependences on progenitor properties and thus provide independent constraints on the stellar models that would be employed to interpret a supernova event.  The clustering of the progenitor types -- three islands corresponding to Wolf-Rayet, BSG, and RSG stars appear in Fig.~\ref{shockprop}  -- shows that the requirements for a successful ``chase" depend greatly on the progenitor.

Although there are variations even within each class, local SBO signals are expected to be quite bright.  Estimating the luminosity using Eq.~(\ref{equc1}) for BSG (or $E_\mathrm{se}$ for $n$\,$=$\,$3/2$ from \citealt{Matzner(1999)} for RSG) along with the above durations yields ranges of $\sim\,$1--5$\,\times10^{45}$~erg~s$^{-1}$ for RSG and $\sim\,$3--10$\,\times10^{44}$~erg~s$^{-1}$ for BSG, which is $\gtrsim\,$$10^{11}\,L_\odot$ for even the least luminous.  Assuming a distance of 5~Mpc leads to energy fluxes in the range $\sim\,$$10^{-7}-$$10^{-6}$~erg~cm$^{-2}$~s$^{-1}$, with RSG temperatures of $\sim\,$3--6$\,\times10^5\,$K and $\sim\,$1--3$\,\times10^6\,$K for BSG.

To determine the shock propagation time (or to have a reasonable probability of doing so) scans must be done on times at least comparable to the much shorter SBO duration.  If the SBO is identified, the determination of the light curve can proceed.  Observations now imply that RSGs are the typical progenitors of Type II-P SNe, which constitute $\sim\,$three out of five core-collapse SNe \citep{Smartt:2008zd}.  RSGs are also the best candidates for successful observation of both SBO arrival time and duration, with the available $10^3-10^4\,$s possibly allowing multiple observations of each candidate galaxy following the neutrino alert.

The likelihood of SBO observation can be increased by combining data from the neutrino burst with information on likely sources within our neighborhood, such as that given in Table~\ref{tab:params}.  In this way telescope time could be allocated to the most likely host galaxies, in proportion to the expectations that these galaxies might host detectable events.  In principle neutrino detection could provide directional information on the host, depending on the number of events detected (as discussed earlier).  

A model for the distribution of core-collapse notices, and general requirements of a telescope useful for SBO searches, can be taken from the remarkably successful {\it Swift} satellite \citep{Gehrels:2004am,Kanner:2012di}.  Upon detection, information will have to be quickly disseminated so that predefined searches can be initiated.  Instruments such as {\it Swift}, with the ability to rapidly slew and observe simultaneously in both the near-UV and soft X-ray, would be invaluable to observe the spectral peak.  For instance, the SBO from the Type Ib SN~2008D was cleanly seen by the {\it Swift} {\it XRT} from a distance of 27~Mpc \citep{Soderberg:2008uh}.  However, the short propagation times and durations for compact BSG/W-R progenitors will likely necessitate an alternative to repeated pointing and slewing of an entire space telescope due to the requisite higher cadence of observations.

One way to get around this constraint would be to utilize wide-field instruments.  A soft X-ray mission (e.g., \citealt{Gehrels2012,Osborne:2013rnr}) could achieve a sensitivity to SBO events from distances much larger than the $\lesssim\,$10~Mpc scale of interest for neutrinos (see \citealt{Calzavara:2003ap} for detailed prospects).  A system of UV space telescopes aiming to reach SBO up to $>\,$100~Mpc is discussed by \citet{Sagiv:2013rma} (see their Fig.~2).  There also exist conceptual studies of rapidly slewing space-bound telescopes that could achieve response times 30--100 times shorter than {\it Swift}, potentially to one second, over substantial ($\pm 35^\circ$) fields of view \citep{Jeong2013}.

Another possibility is to instead try to catch the longer-wavelength tail.  \citet{Adams:2013ana} discuss a dedicated system that requires only small, ground-based telescopes, estimating that upwards of 35\% of BSG events could be detected at a modest cost in a blind search.  Optimizing such a survey instead for RSG can yield close to three out of four such events.  Additionally, the calculations of \citet{Lovegrove:2013ssa} and \citet{Piro:2013voa} indicate that if core-collapse proceeds to a black hole, long, relatively-cool optical transients may follow.  \citet{Piro:2013voa} suggests a $\sim\!10^4\,$K SBO lasting from 3--10 days, thus amenable to such ground-based searches.

A coordination of observational techniques is needed to handle the wide variety of environments that a SN can occur within.  For example, our galaxy sample indicates that NGC~253 and M82 are among the most likely SN hosts, yet both have highly-extincted starburst regions.  The effects of intervening absorption will vary depending on the band observed (see, e.g., \citealt{Calzavara:2003ap}).  Additionally, it should be determined in advance what each instrument that will be available can contribute to such searches.  For instance, although a single slewing space telescope might not feasibly conduct a comprehensive survey for short duration events alone, it would not occupy much time to observe one or a few of the most likely host galaxies guided by the neutrino count data.  Such an approach must also account for the time evolution of the SBO spectrum due to cooling (as illustrated in Fig.~1 of \citealt{Sagiv:2013rma}).  Ideally, the process of partitioning search strategies would be automated.

\section{The Aftermath}
Knowing that the SBO is on the way would also allow for rapid observations of the ensuing optical SN, so that these measurements can be made at the earliest possible times (e.g., \citealt{Stritzinger:2002kg,Quimby,Ofek:2010}).  This will be useful for studying the SN and the star, e.g., by inferring the energy imparted by the shock \citep{Arnett:1996ev}.  If no SBO is found, prompt observation of the SN would be important as a bound on the shock propagation time, which then could lead to an estimate of the probability the SBO was missed, using qualitative correlations such as those in Fig.~\ref{shockprop}.

Observations of the SN progenitor by surveys of supergiants in the local universe (e.g., \citealt{Kochanek:2008mp}) would imply a stellar mass, although models are needed to interpret such data.  Thus, the direct determinations of stellar properties afforded by SBO measurements will be valuable.  This is particularly true for cases where progenitor information is ambiguous due to a nearby cluster or binary companion.  One can even imagine taking images during the period between the neutrino warning and the SBO to compare with earlier epochs.

The primary objective discussed here is based on measuring the times of both core collapse and shock emergence.  If surveys are conducted to observe SBO events without a trigger, it is worth considering what neutrino signals would be useful if detection is not independently possible, i.e., looking after the fact for fewer counts than the two for Hyper-Kamiokande or three for 5~Mton used above.  For a Hyper-K size detector, \citet{Ando:2005ka} determined that searching for even single events from the period preceding a SN could be feasible, depending upon the expected background in the time window that must be considered.  Comparing to Fig.~\ref{shockprop}, for SN~Ibc or BSG events, this would be rather short if the SBO is detected and, since the neutrino burst is much briefer, one neutrino event would be sufficient to time the shock passage.  For RSG, comparing the propagation time differences in Fig.~\ref{shockprop} between adjacent progenitor masses, even while keeping energy injection fixed, the CC--SBO delay is most likely $\sim\,$days.  One might try to further model the SBO and attempt to deduce the core collapse timing, although this is what one would like to test.

Using the rate given in \citet{Kistler:2008us}, in a 5~Mton detector the single event background is $\sim\,$2~hour$^{-1}$, complicating the interpretation of a lone detected event for even BSG progenitors.  However, the rate of random double coincidences (within 10~sec) is $\sim\,$0.2~day$^{-1}$, which, while being too high to allow independent triggers, could be low enough for searches of data taken prior to even RSG SBO events.  Using the same core-collapse rate estimators as before, this would add $\sim\,$2--4 events per decade.  This assumes the detector discussed in \citet{Kistler:2008us}.  A multi-megaton detector that utilizes similar detection principles as IceCube, but with much denser instrumentation to reach a $\sim\,$10~MeV threshold and situated deep within IceCube to use the outer detector as a muon veto, has also been put forward \citep{IceCube:2011ac}.  A study of potential designs indicates an effective detector mass upwards of $\sim\,$7.5~Mton could be reached if photomultiplier tube noise can be sufficiently reduced \citep{Boser:2013oaa}, the feasibility of which is being investigated \citep{Schulte:2013dza}.  The count thresholds for this technique will need to be determined for each design, but aim to achieve detection rates comparable to those presented above.

\section{Discussion \& Conclusions}
The possibility of unifying a variety of observational techniques to study the complete chain of events following a core collapse is a difficult, yet compelling, challenge.  Neutrino and gravitational wave telescopes monitoring the cores of massive stars for signs of collapse can supply advanced notice that a shock should have just formed and is on its way to the star's surface.  Neutrinos are promising because (1) from SN~1987A, we understand what to look for and (2) neutrino telescopes can be scaled to continuously cover out to several Mpc.  This will allow concerted searches for the shock breakout, which, despite being thought to be ubiquitous in core-collapse supernovae, perhaps even for those forming black holes often considered as failures, is one of the rarest observed astronomical events due its transient nature and otherwise unpredictable appearance.

We can extract physics from the SBO only if we identify the photons from the collapsed star before they pass us by.  For a Galactic SN, Super-Kamiokande \citep{Ikeda:2007sa} and IceCube \citep{Abbasi:2011ss} can presently alert us to the event, and the pointing capabilities of Super-Kamiokande will limit the SBO search to a cone of $\sim 5^\circ$.  We thus have a fighting chance of detecting the SBO, though the wait to the next SN may be a long one.  Because Galactic SNe only occur perhaps once per human lifetime, it would be tragic if such an event were to occur while we were not fully prepared to conduct the optimal SBO search.  We will discuss such Galactic searches in more detail elsewhere, considering cases where the SN may be either heavily extincted or exceptionally bright, requiring/allowing novel observational techniques.

The most likely next step in distant SN searches, a Hyper-Kamiokande-class neutrino telescope, could provide $\sim\,$1--2 warnings per decade, giving us an order of magnitude more opportunities to identify the SBO.  It is quite possible that several detectors of this class could be constructed in Asia, Europe, and the U.S.\ within the next $\sim\,$ten years, opening up opportunities through collaborations like SNEWS to coordinate observations and combine data sets.  We have seen, as exhibited in Table~\ref{tab:params}, that a limited number of galaxies are of the greatest interest in this endeavor.  We thus emphasize the importance of understanding the star formation rates and distances of galaxies such as NGC 253 and IC~342 as well as possible.

Measuring both the neutrino (or GW) burst and the SBO is the only way to time the shock's passage through the mantle.  We have described how this observable and three other SBO properties are related to progenitor and explosion parameters in simple polytropes.   Were an SBO observed, these connections would, of course, be made in the context of the best available SN and progenitor models.  Interest in short time domain astronomy has spread rapidly (e.g., \citealt{Paczynski:2000at,Kasliwal:2010cu,Djorgovski:2011iy}) and early time SN data has led to a recent revival in theoretical shock breakout interest (e.g., \citealt{Nakar2010,Tominaga2011,Katz2012,Frey,Ro:2013lfa}).  In the interim leading to large neutrino telescope operations, novel observational programs will hopefully break new empirical ground (e.g., \citealt{Abell:2009aa,Kulkarni:2012kz,Adams:2013ana}), while computational advances should allow for multi-dimensional simulations of SN progenitors that include the frenetic last few years of evolution \citep{Arnett2011}, during which changes in the envelope can be rapid and significant (e.g., \citealt{Quataert2012,Smith:2013wia}).

The protocols for acting on the triggers provided by large neutrino detectors should be carefully developed before the next event.  As new steps in technology are taken, such as Hyper-Kamiokande, the development of such protocols should be viewed as an element of the construction plan -- one of the items that should be ready when the detector first turns on.  The same principles can be applied to gravitational waves if advanced detectors are built to sufficient sensitivity to see extragalactic SNe (e.g., \citealt{Sathyaprakash:2011bh}).  The development of modern gamma-ray burst searches with {\it Swift} provides a model for how this can be done and illustrates how a field can be transformed with rapid follow-up observations.

The identification of SBO events will likely prove to be a powerful tool for determining properties of the progenitors and thus correlating progenitor properties with other observables, optical, neutrino, and gravitational.  This progenitor specificity will establish a tighter connection between observation and modeling, providing a better test of our understanding of the supernova mechanism.

We thank Baha Balantekin, John Beacom, Boris Kayser, Chris Kochanek, Chris Matzner, Chris McKee, Casey Meakin, 
Kate Scholberg, Kris Stanek, Todd Thompson, Terry Walker, and Lincoln Wolfenstein for discussions.
MDK acknowledges support provided by NASA through the Einstein Fellowship Program, grant PF0-110074,
WCH by the US DOE under contracts DE-SC00046548 (UC Berkeley) and DE-AC02-98CH10886 (LBNL) and by the Alexander von Humboldt Foundation,
and HY by the LANL LDRD program.

\end{document}